\documentclass[prb,twocolumn,floatfix]{revtex4}
\usepackage{graphicx}
\usepackage{color}

\begin{document}
\title{
Magnetic phases in the correlated Kondo-lattice model
}
\author{Robert Peters}
\affiliation{Institute for Theoretical Physics, University of G\"ottingen, 
Friedrich-Hund-Platz 1,
37077 G\"ottingen, Germany}
\author{Thomas Pruschke}
\affiliation{Institute for Theoretical Physics, University of G\"ottingen, 
Friedrich-Hund-Platz 1,
37077 G\"ottingen, Germany}

\begin{abstract}
 We study magnetic ordering of an extended Kondo-lattice model
 including an additional on-site
 Coulomb interaction between the itinerant states. The model is solved
 in the dynamical mean-field theory using Wilson's numerical renormalization
 group approach as impurity solver. For a bipartite lattice we find 
 at half filling the expected antiferromagnetic phase. Upon
 doping this phase is gradually suppressed and hints towards phase separation
 are observed. For large doping the model exhibits ferromagnetism, the appearance
 of which can at first sight be explained by Rudermann-Kittel-Kasuya-Yosida
 interaction. However, for large values of
 the Kondo coupling $J$ significant differences to a simple Rudermann-Kittel-Kasuya-Yosida picture can be found.
 We furthermore observe signs of quantum critical points for antiferromagnetic
 Kondo coupling between the local spins and band states.
\end{abstract}
\maketitle              

\section{Introduction}
The Kondo-lattice model (KLM) is one of the most analyzed models in
solid state theory due to its large variety of applications. 
Quite generally, the KLM describes the interaction between localized spins and
a band of conduction electrons. One particular class of systems, where such
a situation can be realized, are the transition metal oxides, which have
received considerable interest due to their rich phase diagram
comprising different types of magnetic and orbital order and even
supercondunctivity.\cite{imada:98} This complexity
stems from the 
interplay between the formation of narrow $3d$-bands leading to a
delocalization of these states on the one hand and the local part of the
Coulomb interaction between the $3d$-electrons tending to localize
them.\cite{imada:98} A particularly interesting example is
La$_{1-x}$Ca$_x$MnO$_3$\cite{salamon:01}. In this cubic 
perovskite the five-fold degenerate $3d$ level is split by crystal field into
three-fold degenerate $t_{2g}$, which have the lower energy, and
two-fold degenerate $e_g$ states. These states have to be filled with
$4-x$ electrons, nominally yielding a metal even for $x=0$. However,
taking into account the local Coulomb interaction, three of these
electrons will occupy the $t_{2g}$-states forming an $S=3/2$ high-spin
state due to
Hund's coupling, which interacts ferromagnetically with the electron
occupying the $e_g$ states. Besides its complicated
phase diagram with a large variety of paramagnetic and
magnetically ordered metallic and insulating phases one finds a
colossal magneto-resistance (CMR).\cite{ramirez:97} 
The physics just described can be covered in great parts by the KLM,
in this context usually called Double Exchange Model (DE).\cite{zener:51,millis:95}

Another class of materials which can be addressed by 
the KLM with ferromagnetic exchange interaction between conduction states and
the localized spin degrees of freedom are magnetic semiconductors or
semi-metals in the series of the rare earth monopnictides and monochalcogenides.\cite{ovchinnikov:91,sharma:05,sharma:06}
Here the focus lies on the magnetic respectively magneto-optic properties of
for example EuS, EuO, GdN or thin films of such systems.\cite{kienert:07}

Last but not least the ferromagnetic KLM can be applied to investigate
the magnetic 
properties of diluted magnetic
semiconductors such as Ga$_{1-x}$Mn$_x$As. In these materials the III-V semiconductor
GaAs becomes ferromagnetic by introducing magnetic ions,\cite{ohno:98,dms_review} such
as Mn. The local spin of the Mn-ions couples ferromagnetically to the states
of the semiconductor, a setup which can be modeled by the KLM. Note, however,
that in these materials disorder effects are expected to play a vital role.\cite{tang:06}

A completely different realization of the KLM starts from the periodic
Anderson model 
(PAM),\cite{vid:04,grenzebach:06,hewson:93} a model for heavy fermion
physics.\cite{stewart:01} Heavy fermion physics manifests itself in a
number of lanthanide and actanide compounds, which have a very large
effective mass in common. The low-temperature physics of these
compounds is determined by a partly filled f shell and hybridization
induced spin flip
scattering between the f and conduction electrons. By a
Schrieffer-Wolff-Transformation\cite{schrieffer:66} the periodic
Anderson model can be mapped onto the KLM with antiferromagnetic exchange
interaction. Note that in these materials one typically expects a 
competition between the heavy fermion physics, driven by the Kondo effect
due to the antiferromagnetic coupling, and the formation of antiferromagentism.\cite{doniach:77}

In most of the approaches one does not include Coulomb interaction among the
conduction band electrons.  Especially for manganites this is
quite likely an insufficient 
approximation, because all the physics takes place in a 
correlated d-band, as explained above. There is no reason to ignore the
local Coulomb correlations in the itinerant subshells, in particular because
estimates of its magnitude typically lead to values of the order or even
larger than the bandwidth of the d states at the Fermi level.\cite{imada:98}
Therefore, we want to address the influence of these local Coulomb 
correlations among the conduction band on the magnetic properties of the
KLM.

Generally, the 5-fold or 14-fold degenerate d- respectively f-shells
split in a crystalline environment into more or less well separated
subshells. Here, we assume that the different crystal field levels are
well separated on the scale of relevant low-energy structures -- this is
typically true in transition metal compounds, but less obvious in rare earth
systems -- and
focus on a situation, where we have one subshell
well localized, hosting a spin $S$, while the remaining
levels are split sufficiently to leave one relevant, spin-degenerate
state close to the Fermi energy. Our model thus consists of
a conventional one-band Hubbard-Model,\cite{hubbard:63,gutzwiller:63,kanamori:63}
where the itinerant states are in addition coupled to  local spins $S$, i.e.
\begin{equation}
H=H_{\rm Hub}+H_{\rm Spin}\;\;. 
\label{model}
\end{equation}
$H_{\rm Hub}$ describes the ordinary Hubbard Model,
\begin{displaymath}
H_{\rm Hub}=-t\sum_{<i,j>,\sigma}\hat{c}_{i,\sigma}^\dagger \hat{c}_{j,\sigma}+U\sum_i \hat{n}_{i,\uparrow}\hat{n}_{i,\downarrow}\;\;,
\end{displaymath}
where $\hat{c}_{i,\sigma}^\dagger(\hat{c}_{i,\sigma})$ creates (annihilates) an
electron at lattice 
site $R_i$ with spin $\sigma$. While the inclusion of an arbitrary hopping 
$t_{ij}$ presents no principle
problem, we focus on the simplest case of next-neighbor-hopping
parametrized by an amplitude $t$. Finally, $\hat{n}_{i,\sigma}=\hat{c}^\dagger_{i,\sigma}\hat{c}_{i,\sigma}$ denotes the
density operator for a $\sigma-$electron at site $R_i$ and $U$ parametrizes
the local Coulomb interaction.

The band electrons are in addition coupled to a local spin $S_i$ at each lattice
site by an exchange interaction
\begin{displaymath}
H_{\rm Spin}=-J\sum_i\vec{s}_i\cdot\vec{S}_i\;\;,
\end{displaymath}
where $\vec{s}_i$ is the spin operator for the band states at site $R_i$.
As discussed before, such a term can arise through Hund coupling, in which case $J$ is
ferromagnetic, or through a hybridization, leading to an
antiferromagnetic $J$.\cite{schrieffer:66} Note that both effects can
appear simultaneously, thus partially compensating each
other.\cite{nekrasov:03}

Even without Coulomb interaction $U$ this model is not exactly
solvable in general, thus approximations have to be made. 
For $U=0$, 
a first approach can consist of a perturbative treatment of the exchange
interaction $J$, leading to the well-known effective Rudermann-Kittel-Kasuya-Yoshida
(RKKY) interaction\cite{rudermann:54,kasuya:56,yosida:57} with a characteristic,
dimension-dependent dependence on distance. 
Although generally accepted as an at least proper ansatz for a qualitative
discussion of magnetic properties of models like the KLM, it has not yet been
studied in detail how well this approximation works for increasing $J$.
Furthermore, for the antiferromagnetic Kondo lattice model, where heavy Fermion
physics can play an essential role, or in the presence of additional correlations
in the band states, the validity of the use of the RKKY arguments is far from
clear. Thus one aspect of the present paper is to investigate to what extent
the RKKY exchange indeed leads to a reasonable description of the low-temperature
properties of the KLM.

As approximation to study the KLM while leaving as much of the local correlations
induced by both the Coulomb interaction $U$ and the exchange $J$ intact as possible
we use the Dynamical
Mean Field Theory (DMFT),\cite{georges:96} mapping the lattice model
onto an effective impurity problem. In former treatments of the model (\ref{model}), especially within
DMFT,\cite{furukawa:94,held:00c,sen:06} classical spins were assumed to avoid
the sign problem of the quantum Monte-Carlo treatment of the effective
impurity problem employed there.  
The importance of a fully quantum mechanical treatment of the local
spins even 
in impurity problems has been addressed by \textcite{peters:06a}
and the effects of quantum spins in the KLM by \textcite{kienert:06}. 
To achieve such a fully quantum mechanical treatment we use
Wilsons's 
Numerical Renormalization Group (NRG)\cite{wilson:75,krish:80a,bulla:07}
as impurity solver. The NRG can handle the
whole interaction and temperature regime and is also able to calculate
Green's function even in ordered phases. We use a recent improved
technique for calculating Green's
functions\cite{anders:05,peters:06b,weichselbaum:06} within NRG. 
Therefore we are able to treat a large bandwidth of values of both on-site
interactions $J$ and $U$. 

The paper is organized as follows. In the next section we discuss the form
of the perturbative RKKY exchange in the limit spatial dimension $D\to\infty$
appropriate for the DMFT. In section III we present our results for the
magnetic phase diagrams of the model (\ref{model})
at half-filling and finite doping and different values of $U$ and
discuss their dependence on the sign and magnitude of $J$. A summary will
conclude the paper.
\section{RKKY-interaction in the limit $D\to\infty$}
Conventionally the interaction between localized spins in metals is analyzed
in terms of the 
RKKY interaction.\cite{rudermann:54,kasuya:56,yosida:57}
This effective exchange interaction shows a characteristic dependence on
distance $R$ and local exchange coupling $J$ of the form
\begin{equation}\label{eq:rkky_typical}
J_{RKKY}\propto J^2\frac{\cos(k_FR)}{(k_FR)^\alpha}\;\;,
\end{equation}
where $k_F$ is the Fermi momentum of the host and $\alpha$ some positive,
dimension-dependent number. This distance-dependence is due to the sharp
Fermi surface in metals and, strictly speaking, valid only in the limit $J\to0$.
Note that, because $J_{RKKY}\propto J^2$, the sign of $J$ does not matter.
Furthermore, it is a priori not evident, how the
distance-dependence looks like in the limit spatial dimension $D\to\infty$,
which sets the framework for the construction of the DMFT. In particular,
in this limit there is no proper definition of $k_F$, which controls the
dependence of $J_{RKKY}$ on the occupancy of the band. Thus, in order to 
be able to study to what extent the RKKY interaction controls the magnetic
properties of the KLM in DMFT, we will derive this interaction and especially
its dependency on the filling for the limit $D\to\infty$ in the following.

To this end
we consider the extended Hubbard model (\ref{model})
on a hypercubic lattice.
In the
paramagnetic phase with $\langle
\vec{S}_i\rangle=\langle\vec{s}_i\rangle=0$, 
an effective Hamiltonian for the local spins can be calculated perturbatively
in $J$ by formally tracing out the band states. The resulting expression
in lowest order in $J$ reads
\begin{eqnarray}\label{eq:energy}
H_{\rm eff} &\approx&
-J^2\sum\limits_{ij}\sum\limits_{\alpha=1}^3
\int\limits_0^\beta d\tau
S_i^\alpha S_j^\alpha\langle s_i^\alpha[\tau] s_j^\alpha[0]\rangle_{J=0}
\nonumber\\
&=& 
-J^2\sum\limits_{ij}\sum\limits_{\alpha=1}^3S_i^\alpha
S_j^\alpha\chi_{ij}^{\alpha\alpha}\nonumber\;\;,
\end{eqnarray}
where $\chi_{ij}^{\alpha\alpha}$ denotes the static susceptibility of the
bare Hubbard model. In the paramagnetic case, $\chi_{ij}^{\alpha\alpha}$
does not depend on $\alpha$, and we are free to evaluate it for $\alpha=3$,
for example. This leads to an effective spin-spin interaction
\begin{eqnarray}\label{eq:jrkky}
J_{\rm RKKY}(R_i-R_j) &=& J^2\chi_{ij}^{zz}\nonumber\\
&=&
J^2\frac{1}{N}\sum\limits_q e^{iq\cdot(R_i-R_j)}\chi^{zz}(q)
\end{eqnarray}
Let us evaluate the latter sum for a model with nearest-neighbor hopping
in the limit $D\to\infty$.
As has been shown by M\"uller-Hartmann,\cite{mueller:89} the $q$-dependence then enters only
via 
$$
\eta(q):=\frac{1}{D}\sum\limits_{l=1}^D\cos(q_la)\;\;,
$$
where $a$ is the lattice parameter. 

Due to inversion symmetry,
we have
\begin{widetext}
$$
\frac{1}{N}\sum\limits_q e^{iq\cdot(R_i-R_j)}\chi^{zz}(q)
=
2\frac{1}{N}\sum\limits_q \cos\left(q\cdot(R_i-R_j)\right)\chi^{zz}(q)\\
=
\int\limits_{-1}^1dx 
\varrho_{ij}(x)
\chi^{zz}(x)
$$
\end{widetext}
with
$$
\varrho_{ij}(x):=
2\frac{1}{N}\sum\limits_q\left[
\cos\left(q\cdot(R_i-R_j)\right)\delta(x-\eta(q))
\right]\;\;.
$$
For a nontrivial result in the limit $D\to\infty$ we need $D\cdot J_{\rm RKKY}$
to be finite. We thus will evaluate
$$
D\int\limits_{-1}^1dx 
\varrho_{ij}(x)
\chi^{zz}(x)
$$
directly. Following again the arguments by M\"uller-Hartmann,
we first rewrite
$$
\delta(x-\eta(q))=\int\limits_{-\infty}^\infty
\frac{ds}{2\pi} e^{i(x-\eta(q))s}
$$
and obtain
\begin{widetext}
$$
D\varrho_{ij}(x)
=
2D
\int\limits_{-\pi}^\pi\frac{d^Dq}{(2\pi)^D}
\int\limits_{-\infty}^\infty\frac{ds}{2\pi}
e^{i(x-\eta(q))s}
\cos\left(q\cdot(R_i-R_j)\right)\;.
$$
\end{widetext}
Expanding the exponential in terms of $\eta(q)$ and observing that
\begin{eqnarray*}
\int\limits_{-\pi}^\pi\frac{d^Dq}{(2\pi)^D}\cos\left(q(R_i-R_j)\right)
&=&\delta_{i,j}\\
\int\limits_{-\pi}^\pi\frac{d^Dq}{(2\pi)^D}\cos\left(q(R_i-R_j)\right)
\sum\limits_{l=1}^D\cos(q_l\cdot a)&=&\delta_{|R_i-R_j|,a}\;\;,
\end{eqnarray*}
and so on for terms involving $\eta(q)^m$ for $m>1$, we obtain for $i\ne j$
\begin{eqnarray}
2D\varrho_{ij}(x)&\stackrel{D\to\infty}{=}&
2\int\limits_{-\pi}^\pi\frac{d^Dq}{(2\pi)^D}
\int\limits_{-\infty}^\infty\frac{ds}{2\pi}
e^{ixs}\left[D-iD\eta(q)s+\ldots
\right]\nonumber\\ & &\cdot
\cos\left(q\cdot(R_i-R_j)\right)\nonumber\\
&=&
-i\delta_{|R_i-R_j|,a}
\int\limits_{-\infty}^\infty\frac{ds}{2\pi}e^{ixs}
s+O\left(\frac{1}{D}\right)\nonumber\\
&=&-\delta_{|R_i-R_j|,a}\,\frac{d}{dx}\delta(x)\,+O\left(\frac{1}{D}\right)\nonumber
\;\;.
\end{eqnarray}
With this result we find in the limit $D\to\infty$ for the RKKY exchange
\begin{equation}\label{eq:rkky_final}
D\cdot J_{\rm RKKY}(R_i-R_j)
=
J^2\,
\delta_{|R_i-R_j|,a}\,
\left.\frac{d}{dx}\chi^{zz}(x)\right|_{x=0}
\end{equation}
Note that the right hand side of (\ref{eq:rkky_final}) is already correctly
scaled to obtain nontrivial results in the limit $D\to\infty$. Furthermore,
the RKKY exchange acts only on nearest neighbors, the oscillatory
structures arising from Fermi surface singularities in finite
dimensions are absent in $D\to\infty$. Nevertheless, the
dominant nearest-neighbor exchange constant which through its
sign determins the
type of order can be obtained. 

\begin{figure}[htb]
\begin{center}
\includegraphics[width=0.45\textwidth,clip]{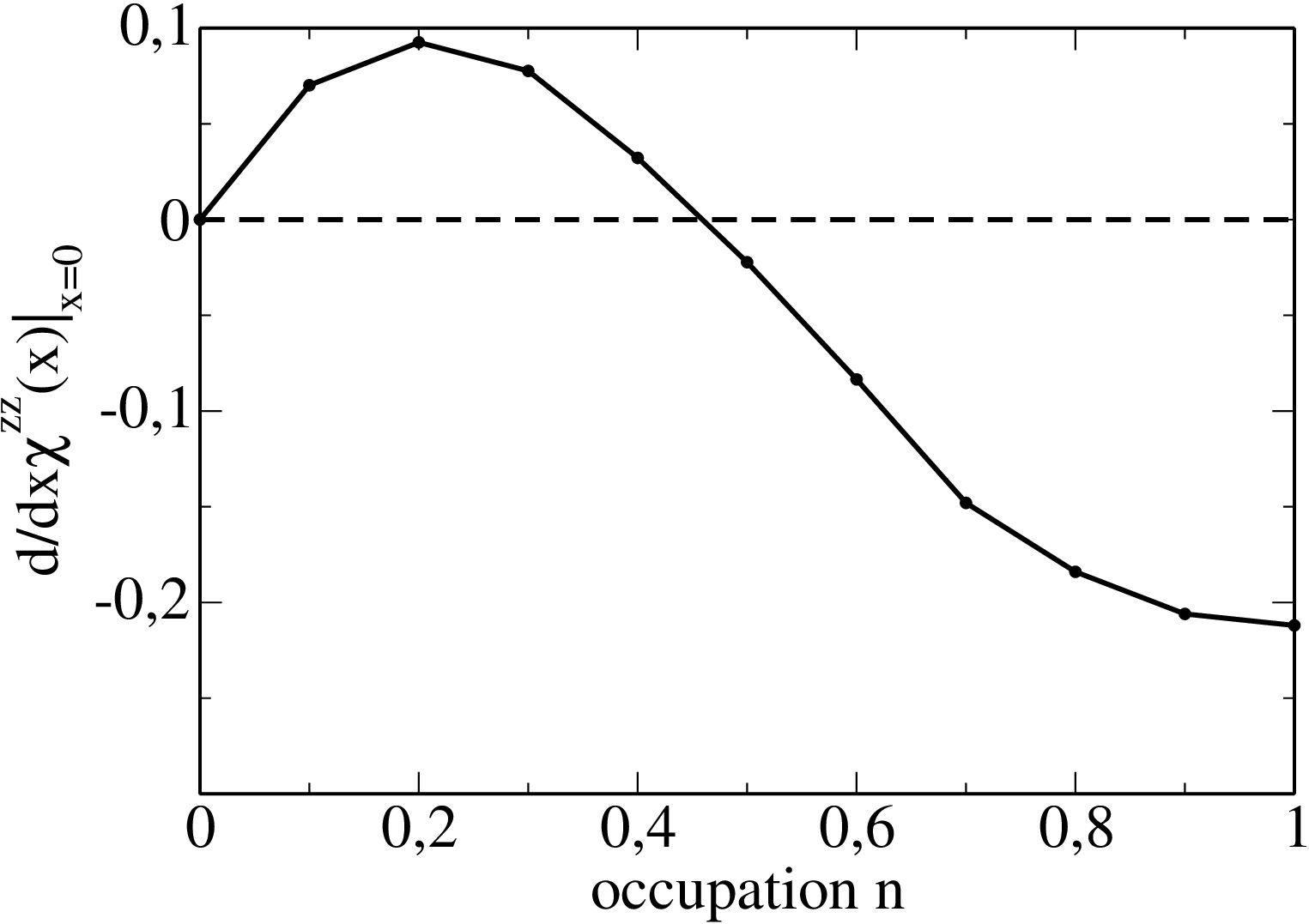}
\end{center}
\caption[]{$\frac{d}{dx}\chi^{zz}(x)\vert_{x=0}$ for $U=0$ as function of
the occupancy $n$}\label{rkkystrength}
\end{figure}

For $U=0$, the susceptibility is given
by the simple bubble, which can be evaluated exactly albeit only numerically,
leading to the behavior for $J_{RKKY}$ depicted in Fig.\ \ref{rkkystrength}. Note
that the sign change from $J<0$ (afm) to $J>0$ (fm) appears for
$n<0.5$. We will discuss this figure in connection with numerical
results in the next sections.

For finite $U>0$, a similar evaluation is not possible, because the susceptibility
entering now is the full lattice susceptibility. While $\chi_{zz}(x=0)$ is the
local susceptibility, which can be obtained from the effective impurity problem
directly, its derivative determining $J_{RKKY}$ involves neighboring $x$
values, which cannot be calculated within NRG. We content ourselves here by
noting that finite Coulomb correlations typically lead to a more
asymmetric distribution of spectral weight, which is known to tend to enhance
ferromagnetic correlations. We thus expect that the root of $J_{RKKY}$ will
shift to larger values of the occupancy $n$ for finite $U$.
\section{Results}
In this section we present our results for the magnetic phases of the extended
KLM (\ref{model}). Up to now the local spin $S$ was completely arbitrary and
can in principle take any value. Although it is surely interesting to study 
the effect of increasing spin quantum number on the results,\cite{kienert:06}
we restrict the discussion to the case $S=1/2$ here. 

The calculations were done for a Bethe lattice employing a bipartite
subdivision to accommodate antiferromagnetic order.\cite{georges:96,pruschke:05}
We used a Bethe lattice instead of a hypercubic one for computational
reasons. However, we did not find any significant differences between
the two lattices in test calculations. As discretization parameter
for the NRG we used $\Lambda=2$ and typically kept $1000\ldots2000$ states
in each NRG step.
As our unit of energy we choose the bandwidth $W=4t$. 
\subsection{Antiferromagnetism at half filling}
Let us begin by examining the magnetic order at half-filling, $n=1$. Clearly
Fig.\ \ref{rkkystrength} states that the interaction between the
localized spins is negative which is supposed to result in
antiferromagnetic order. 
\begin{figure}[htb]
\begin{center}
\includegraphics[width=0.45\textwidth,clip]{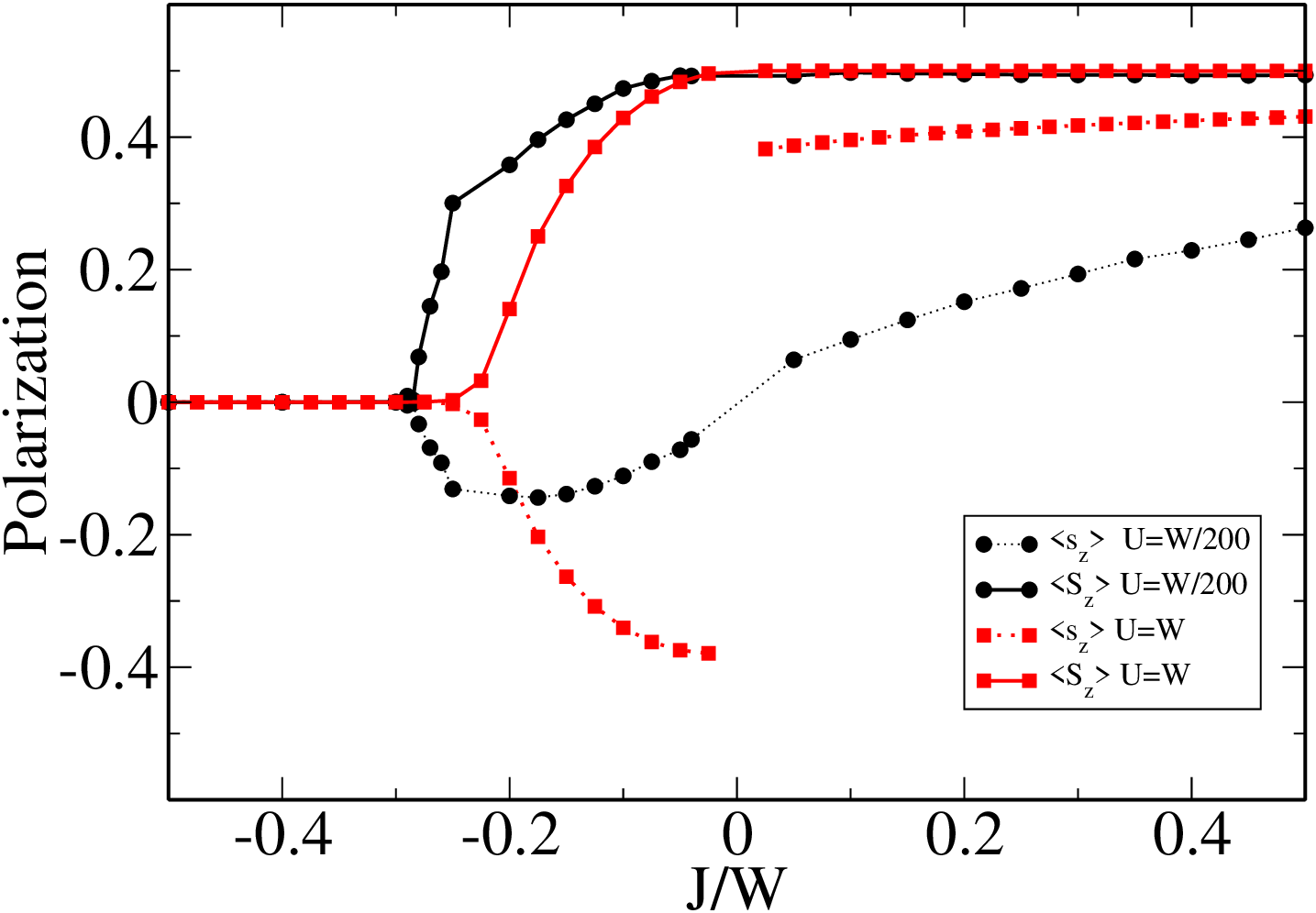}
\end{center}
\caption[]{(color online) Antiferromagnetic polarization
  at half-filling and $T=0$ as function of the coupling $J$ between spins and
  band electrons. Circles were calculated for $U/W=1/200$,
  corresponding approximately to $U=0$, 
  squares for $U=W$.}\label{antihalf}
\end{figure}
In Fig.\ \ref{antihalf} are collected results for the polarization
$\langle s_z\rangle=\frac{1}{2}(n_\uparrow-n_\downarrow)$ of the band electrons
and $\langle S_z\rangle$ of the local spin.
One can see in
Fig.\ \ref{antihalf} two curves corresponding 
to $U\approx 0$ (circles) and $U=W$ (squares). For large negative $J$ there
is no magnetic order. The system forms a Kondo-insulator, locally quenching
all moments. Within DMFT we find for the critical value of the
coupling at $U=0$ $J_c/t\approx-0.3W/t= -1.2$ which is consistent with long-known results.\cite{lacroix:79} With increasing $U$, this value is shifted to somewhat smaller absolute
values $|J|$. 

For $J>J_c$ antiferromagnetic order can emerge
before all 
moments have been quenched locally. Quite generally, we observe that the
polarization $\langle s_z\rangle$ of the band states and $\langle S_z\rangle$
of the local spins are opposite in sign for $J<0$. Around the critical
value $J_c$ the resulting total polarization $\langle s_z+S_z\rangle$ thus
is very small, although both contributions can already have rather sizable
values. This result is quite interesting, in particular in view of the so-called
small-moment antiferromagnetism observed in several rare-earth compounds.

Obviously, for $U=0$ there is no
antiferromagnetic order at $J=0$; however, even for very small $\vert
J/W\vert\ll 1$ the local spins are almost fully polarized, $\langle S_z\rangle\approx
0.5$. On the other hand, the polarization of
the conduction electrons $\langle s_z\rangle$ goes smoothly through zero.
We can thus identify this range of $J$ values as corresponding to the RKKY
regime, where the local spins are fully polarized and the band electrons show
a polarization proportional to the ``effective field'' $\sim J\langle S_z\rangle$ 
provided by the local spins.
In this region we also observe that the magnetic properties of the system are roughly
independent of the sign of $J$, as predicted by RKKY. 

On the other hand, the behavior in the vicinity of $J_c$ cannot be understood
in terms of RKKY any more, although $|J_c|=0.3W$ is still significantly smaller than $W$ and one might expect the
perturbational arguments to be still valid. However, the physical properties are radically
different for $J<0$ and $J>0$, as no critical point exists for $J>0$ and the
model always is in the fully polarized state.

For
$U>0$ there is antiferromagnetic order even at $J=0$, which represents the pure
Hubbard model.\cite{georges:96,pruschke:05} As mentioned before, for $J<0$ spin
and electron polarizations are antiparallel while for $J>0$ both
orientations are parallel.
Due to to numerical errors we were not able to determine the
behavior as $J\rightarrow 0$. Even with as many as $4000$ states kept in
each NRG step DMFT+NRG calculations
did not converge but showed strong fluctuation for $\vert
J/W\vert<0.04$. The results obtained nevertheless suggest a jump at $J=0$.
Again, as $J\to0^-$, the net polarization $\langle s_z+S_z\rangle$ is strongly
reduced from the almost full values for each individual part.

\begin{figure}[htb]
\begin{center}
\includegraphics[width=0.45\textwidth,clip]{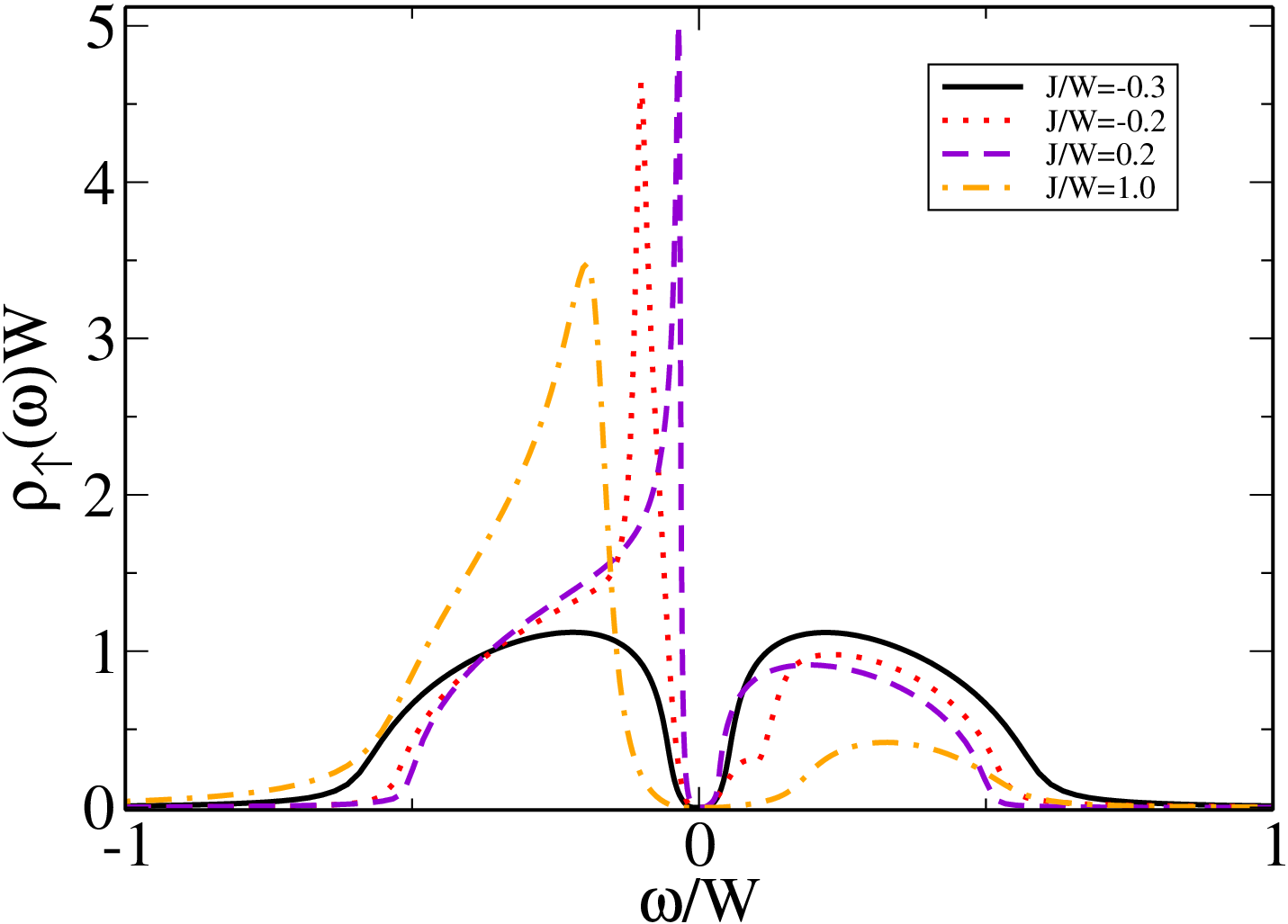}
\end{center}
\caption[]{(color online) Spectral function for the majority spin at half filling
  for $U/W=1/200$ and temperature $T/W=2.3\cdot10^{-4}$.}\label{spekhalf} 
\end{figure}
Fig.\ \ref{spekhalf} shows the majority spectral functions for
$U/W=1/200$ and different $J$. All calculations show a gap at
$\omega=0$, characteristic for an  insulating behavior. The solid
line represents the results for a large antiferromagnetic coupling, i.e.\ in
the Kondo insulating state. The
spectral function agrees with the typical result for paramagnetic phase of a
Kondo-insulator. The remaining 
lines correspond to $J>J_c$. The dotted and dashed lines correspond to
the same value $\vert J\vert$, however with opposite signs. The general shape
is that of a weak-coupling antiferromagnet\cite{zitzler:02} with the
characteristic square-root van-Hove singularities at the gap edges. Note that
for $J<0$ we find a shallow shoulder for $\omega>0$ which quite likely is
the remnant of a quasiparticle peak due to the Kondo effect expected in this regime.\cite{peters:06a} 
This is again a sign, that already for moderate values of $J$ the physical
properties are significantly different for $J<0$ and $J>0$, i.e.\ cannot be
understood by pure RKKY physics.

The dotted-dashed line in Fig.\ \ref{spekhalf} corresponds to a large
ferromagnetic coupling. Here, life-time effects due to a large self-energy
already substantially dominate the spectrum.
\subsection{Doped model at $U=0$}
\begin{figure}[htb]
\begin{center}
\includegraphics[width=0.45\textwidth,clip]{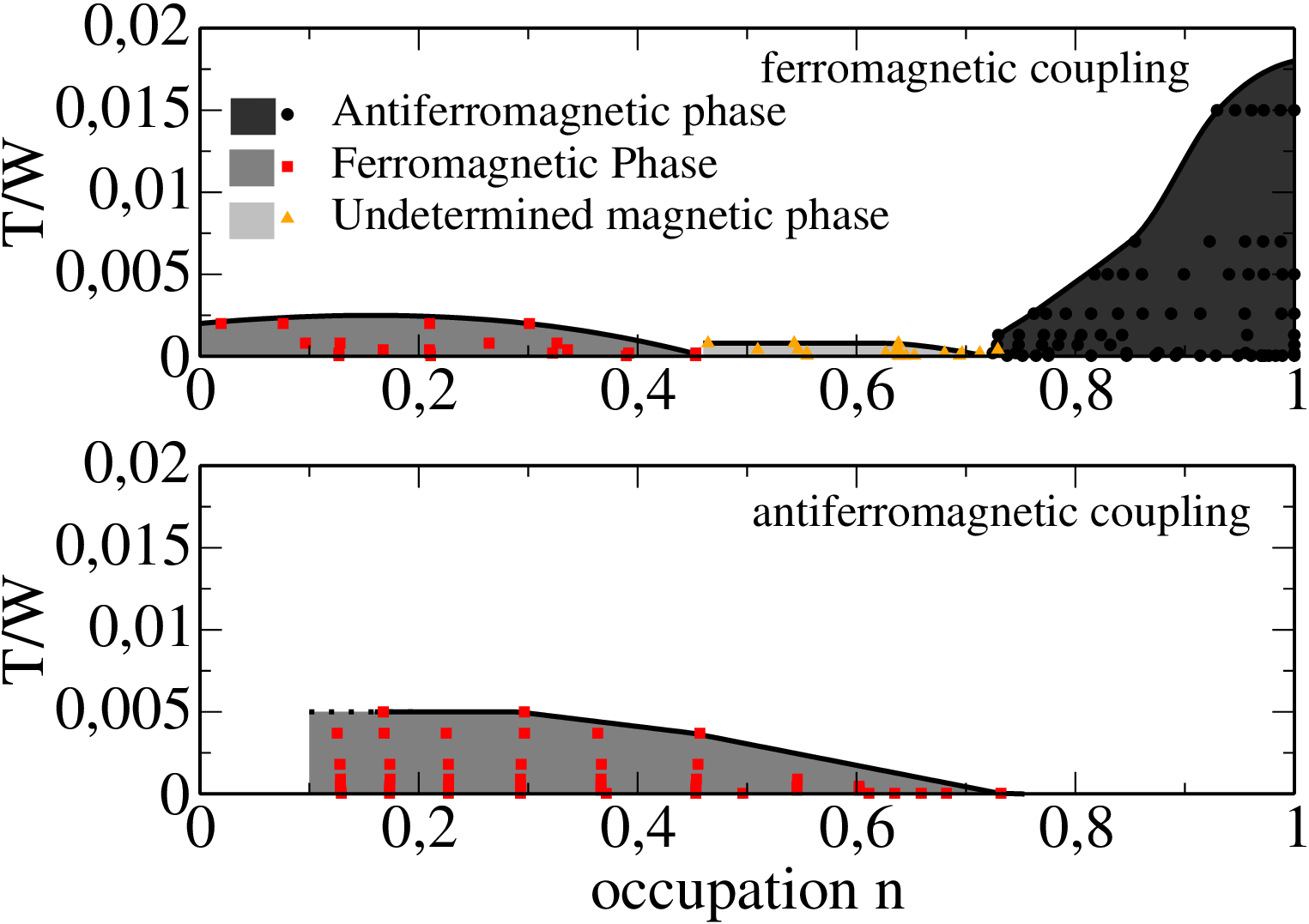}
\end{center}
\caption[]{(color online) Magnetic phase diagrams for $U=0$ and $\vert J\vert=W/2$ as function
  of filling and temperature. The upper (lower) panel shows the results for
  ferromagnetic (antiferromagnetic) coupling. Lines are meant as guide
  to the eyes. For ferromagnetic coupling a magnetic phase exists
  between $0.45<n<0.7$, whose nature could not be determined (see text).
  The number of states kept during NRG calculation was increased at phase
  boundaries.}\label{Phase001}
\end{figure}
Figure \ref{Phase001} shows the magnetic phases of the Kondo-lattice
model at $U=0$ as function of filling $n$ and temperature calculated within
DMFT for a Kondo coupling $|J|=W/2$. Note that this value is already larger
than the critical value $J_c$ at half filling, thus we expect that for $J<0$
the physics is governed by local quenching of the moments close to half
filling.

The upper panel in Fig.\ \ref{Phase001} shows the model with ferromagnetic coupling, while the
lower one  displays the results for antiferromagnetic coupling. 
In the ferromagnetic Kondo-lattice model we find an antiferromagnetic
phase around half-filling $n>0.7$, and a 
ferromagnetic phase for $n<0.45$. The location of phase boundary of the
ferromagnet at $T=0$ appears to agree very well with RKKY prediction from Fig.\
\ref{rkkystrength}, where the coupling between the localized spins
within DMFT becomes ferromagnetic for $n<0.45$. We have focused
in our work on homogeneous ferromagnetic-, antiferromagnetic N\'{e}el- and
paramagnetic states. Between $0.45<n<0.7$ calculations showed also
significant magnetic response. But neither a ferromagnetic nor an
antiferromagnetic N\'{e}elstate could be stabilized. The real
nature of this phase could not yet be clarified in our calculations,
but one might argue that one should expect some type of incommensurate order.
In fact, the magnetic
phase diagram of the Kondo-lattice model has been
analyzed by a number of other
authors.\cite{lacroix:79,fazekas:91,yunoki:98,dagotto:98,nagai:99,santos:02,nolting:03b,yin:03,kienert:06,fishman:06}
In addition to the conventional homogeneous ferromagnet and N\'eel antiferromagnet
other magnetic phases like incommensurate, chiral and short range
ordered phases were analyzed for $0.5<n<1.0$ in these studies
and found to be stable in the regime where our calculations fail to converge.

\begin{figure}[htb]
\begin{center}
\includegraphics[width=0.45\textwidth,clip]{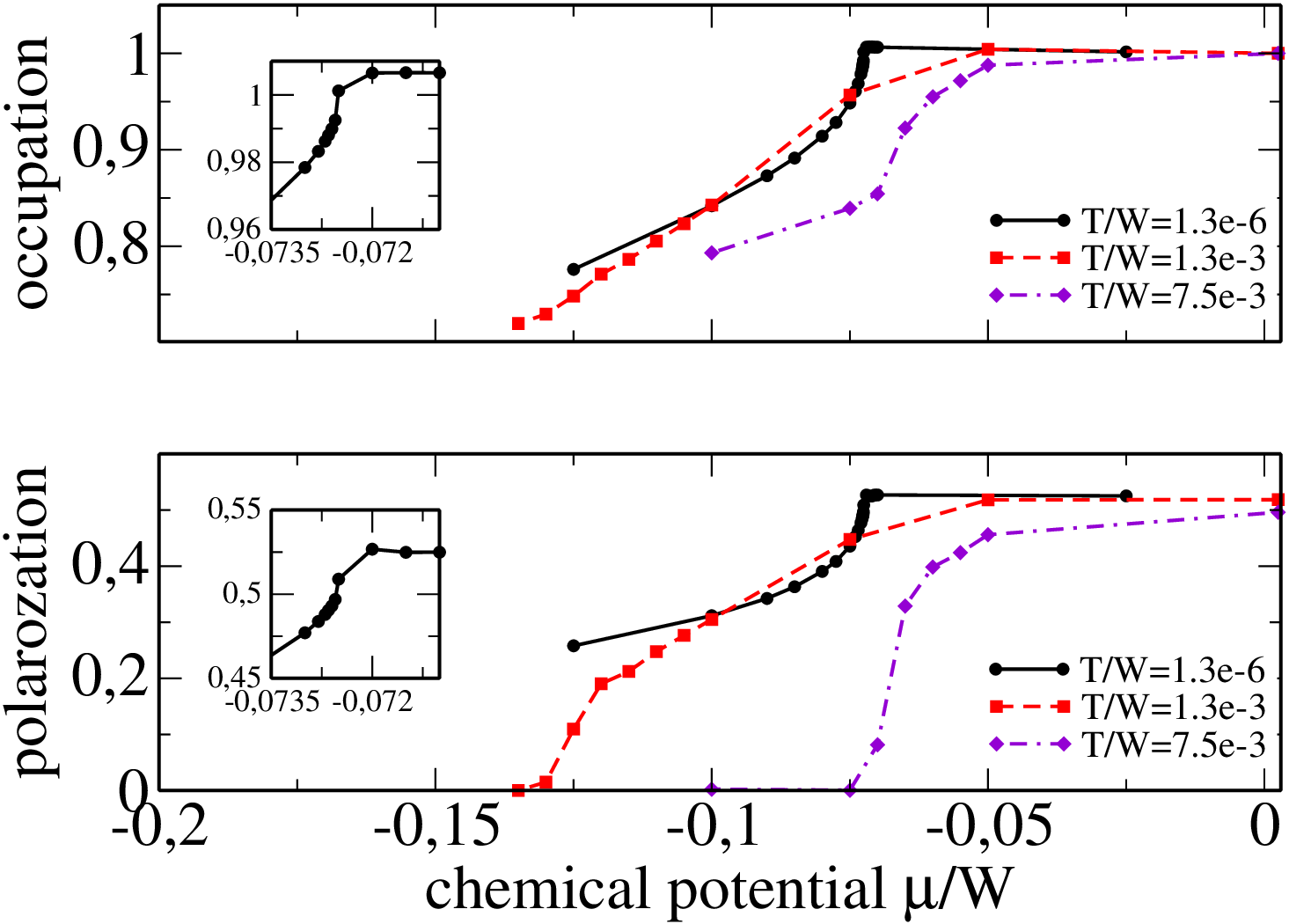}
\end{center}
\caption[]{(color online) Occupancy (upper panel) and polarization (lower panel) for the antiferromagnetic
  solution around half filling as function of chemical potential $\mu$
for different temperatures. Other
parameters as in Fig.\ \ref{Phase001}.}\label{Antiferro001}
\end{figure}
For the antiferromagnet close to $n=1$, several authors\cite{yunoki:98,dagotto:98,nagai:99,yin:03,kienert:06} also reported phase separation between
antiferromagnetism at half-filling and phases away from
half-filling. Figure \ref{Antiferro001} shows our results for the occupancy
and polarization in the antiferromagnetic phase
as function of the chemical potential $\mu$ for different temperatures. We see
that with
decreasing temperature the slope of the curves for $n\to1.0$ get larger
and larger.
However, even for the lowest temperature we could
stabilize every occupation number $n<1.0$, even if there is a large
slope for $n\to1.0$. Increasing the numerical accuracy by for example keeping
more states per NRG step also tend to increase the slope further. However, we
did never observe a clear jump of the occupation at a critical value of the chemical
potential. Note, however, that we cannot rule out phase separation at $T=0$
from our numerical results.

The lower panel in figure \ref{Phase001} shows the phase diagram of
the antiferromagnetic Kondo-lattice model. Here our results yield only
evidence for a ferromagnetic phase. In contrast to the
ferromagnetic Kondo-lattice model, this phase reaches to significantly higher
occupations and temperatures; the antiferromagnetic coupling seems to
stabilize ferromagnetic order. This behavior as well as the rather large
critical value for $n$ at $T=0$ cannot be explained within the RKKY picture.
Furthermore, the quite likely incommensurate
phase neighboring the ferromagnet for $J>0$ is absent here. For
$n\to0$ we could not stablilize a magnetic phase any more due to numerical
problems connected with the small filling.

\begin{figure}[htb]
\begin{center}
\includegraphics[width=0.45\textwidth,clip]{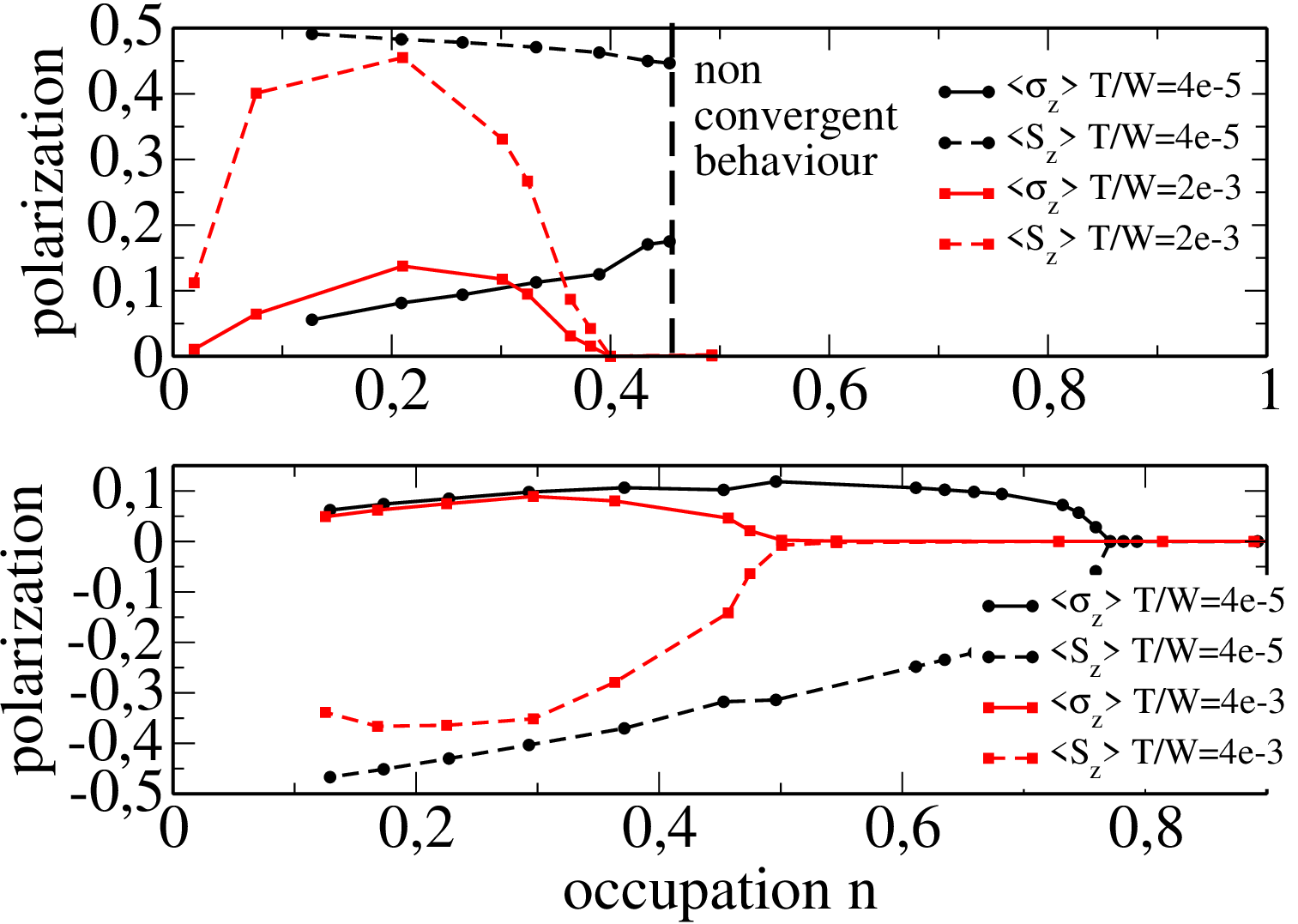}
\end{center}
\caption[]{(color online) Polarization of the ferromagnetic solutions. The upper (lower) panel
  corresponds to ferromagnetic (antiferromagnetic) coupling. Other
parameters as in Fig.\ \ref{Phase001}.}\label{Ferro001} 
\end{figure}
The previous observations are further substantiated by the results in
Fig.\ \ref{Ferro001}, where we show the ferromagnetic polarizations for
ferromagnetic (upper panel) and antiferromagnetic (lower panel)
coupling parameter. The circles correspond to a very low temperature, squares
 to a temperature near the transition into the paramagnetic
state. Full lines represent $\langle s_z\rangle$, dashed lines $\langle S_z\rangle$.
For $J>0$ and temperatures above the critical temperature for the ``incommensurate''
phase, both $\langle s_z\rangle$ and $\langle S_z\rangle$ show a similar
behavior as function of doping. In particular, they vanish smoothly as 
$n\to n_c$, indicating a second order phase transition. For very low $T$,
on the other hand, $\langle S_z\rangle$ remains almost constant at the
fully polarized value until we reach the phase boundary to the ``incommensurate''
phase at $n_c\approx0.45$. At the same time, $\langle s_z\rangle$ monotonically
increases with filling up to $n_c$. Although we cannot determine the true
structure of the phase for $n>n_c$, its existence seems to be connected to
the peculiar behavior of $\langle s_z\rangle$ rather than with $\langle S_z\rangle$,
i.e.\ is primarily driven by the itinerant electrons. We are currently 
further investigating this phase transition by including magnetic structures
other than the N\'eel state in our calculations.

For antiferromagnetic coupling (lower panel of Fig.\ \ref{Ferro001}) we find a
critical $n_c$ increasing with temperature, where the ferromagnetic metal
becomes a Kondo insulator. Note that for $T/W=4\cdot10^{-5}$ we are
effectively in the ground state. Further lowering the temperature does
not change the phase diagram any more. Both $\langle s_z\rangle$ and $\langle S_z\rangle$
appear to vanish smoothly, too, thus again indicating a second order phase transition. 

Choosing $|J|<|J_c|$ does not modify the general structure
of the phase diagram for ferromagnetic coupling. The results
for antiferromagnetic coupling, of course, do change, in particular close
to half filling, where now an antiferromagnetic phase appears, too.
While the overall structure of the phase diagram is now
similar to the case $J>0$ in Fig. \ref{Phase001} upper panel, the values for $n_c^{(FM)}$ are still enhanced
with respect to the RKKY prediction.
\subsection{Finite Coulomb interaction $U=W/2$}
\begin{figure}[htb]
\begin{center}
\includegraphics[width=0.45\textwidth,clip]{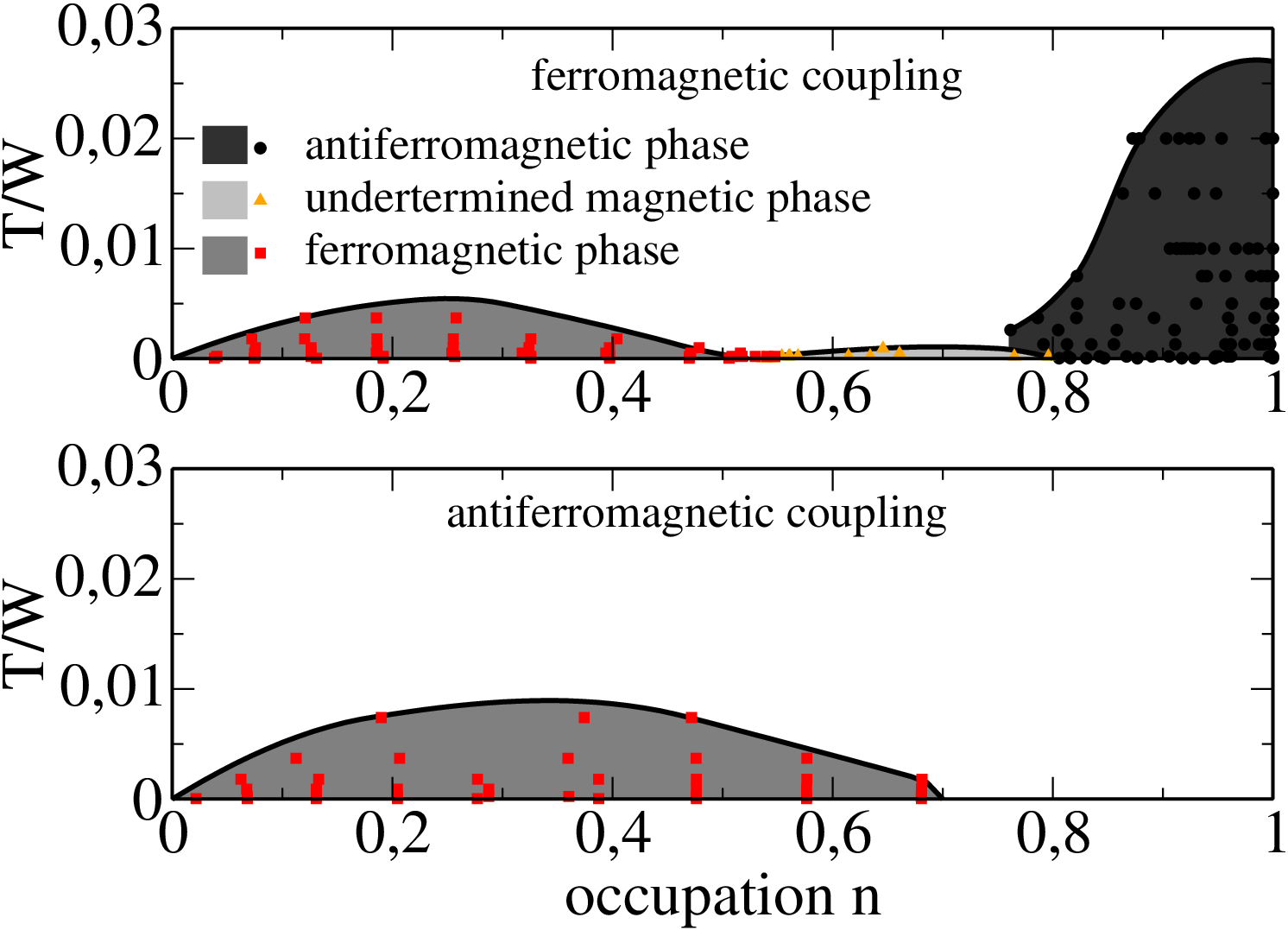}
\end{center}
\caption[]{(color online) Magnetic phase diagrams for $U=W/2$ and $\vert J\vert=W/2$
  as function of filling and temperature. The upper (lower) panel shows results
  for ferromagnetic (antiferromagnetic) coupling. Lines are meant as guides
  to the eye.  For ferromagnetic coupling a magnetic phase exists
  between $0.5<n<0.8$, whose nature could not be determined (see text). The
  number of states kept during NRG calculation was increased at phase
  boundaries.}\label{Phase0101}
\end{figure}
Let us now turn to the model with finite on-site
Coulomb interaction. We present results for fixed $U=\vert J\vert=W/2$,
changing these parameters modifies the observations quantitatively but not
qualitatively.
The phase diagrams for
ferromagnetic (upper panel) and antiferomagnetic (lower panel)
coupling are
shown in Fig.\ \ref{Phase0101}. As for $U=0$, we find both an
antiferromagnetic phase around half filling and a ferromagnetic phase for
larger doping in the ferromagnetic Kondo-lattice model. The critical values
have changed to $n_c^{(AF)}\approx0.8$ and 
$n_c^{(FM)}\approx0.53$ for $T=0$, indicating that local correlations
additionally stabilize ferromagnetic order but destabilize antiferromagnetic
for $J>0$, in accordance with the anticipated variation of $J_{RKKY}$ with $U$.

For the ferromagnetic Kondo lattice model we again find between the
ferromagnetic phase for $n<0.53$ and the antiferromagnetic for $0.8<n<1$ 
a magnetic phase whose nature could not be determined for the same reasons as for
$U=0$.
Guided by our observations for $U=0$ and a comparison to other studies by
other groups, one can again assume the appearance of an incommensurate magnetic
phase like a spin wave or a chiral phase. From quantum Monte-Carlo studies
\cite{freericks:95} for the conventional Hubbard model an 
incommensurate phase can indeed be anticipated in this parameter regime.

\begin{figure}[htb]
\begin{center}
\includegraphics[width=0.45\textwidth,clip]{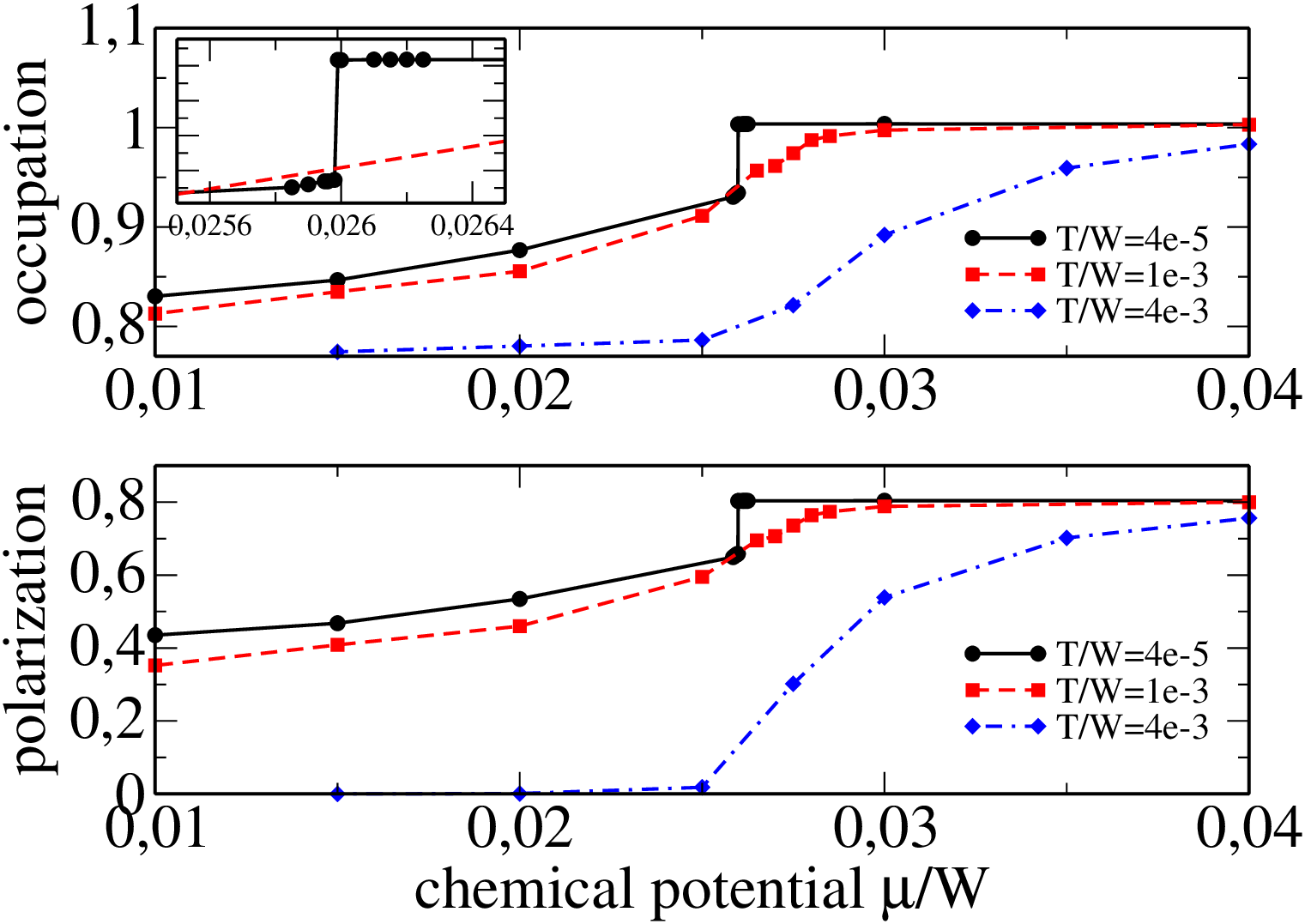}
\end{center}
\caption[]{(color online) Occupancy (upper panel) and polarization (lower panel) for the antiferromagnetic solution
  around half filling as function of chemical potential $\mu$ for different
  temperatures. 
  chemical potential $\mu$. Other parameters as in Fig.\ \ref{Phase0101}.}\label{antisepa}
\end{figure}
Figure \ref{antisepa} shows occupation and
polarization as function of the chemical potential $\mu$ for different
temperatures. For high-temperatures, squares and triangles, there is
a smooth crossover from half-filling to lower occupancies, as for
$U=0$ (cf.\ Fig.\ \ref{Antiferro001}).
However, in contrast to the indecisive results at $U=0$, we observe clear
evidence for phase separation at $T/W=4\cdot 10^{-5}$, circles, this time. 
There exists a critical value $\mu_{crit}$, where both occupation and
polarization jump from their values at half-filling to a 
smaller occupancy and polarization. We therefore have phase separation between
the antiferromagnetic insulator at half-filling and an
antiferromagnetic metal with a lower occupancy. Note that one also finds phase
separation in the conventional Hubbard model, which here however occurs between
an antiferromagnetic insulator with $n=1$ and a paramagnet with $n<n_c^{(AF)}$.\cite{zitzler:02}

\begin{figure}[htb]
\begin{center}
\includegraphics[width=0.45\textwidth,clip]{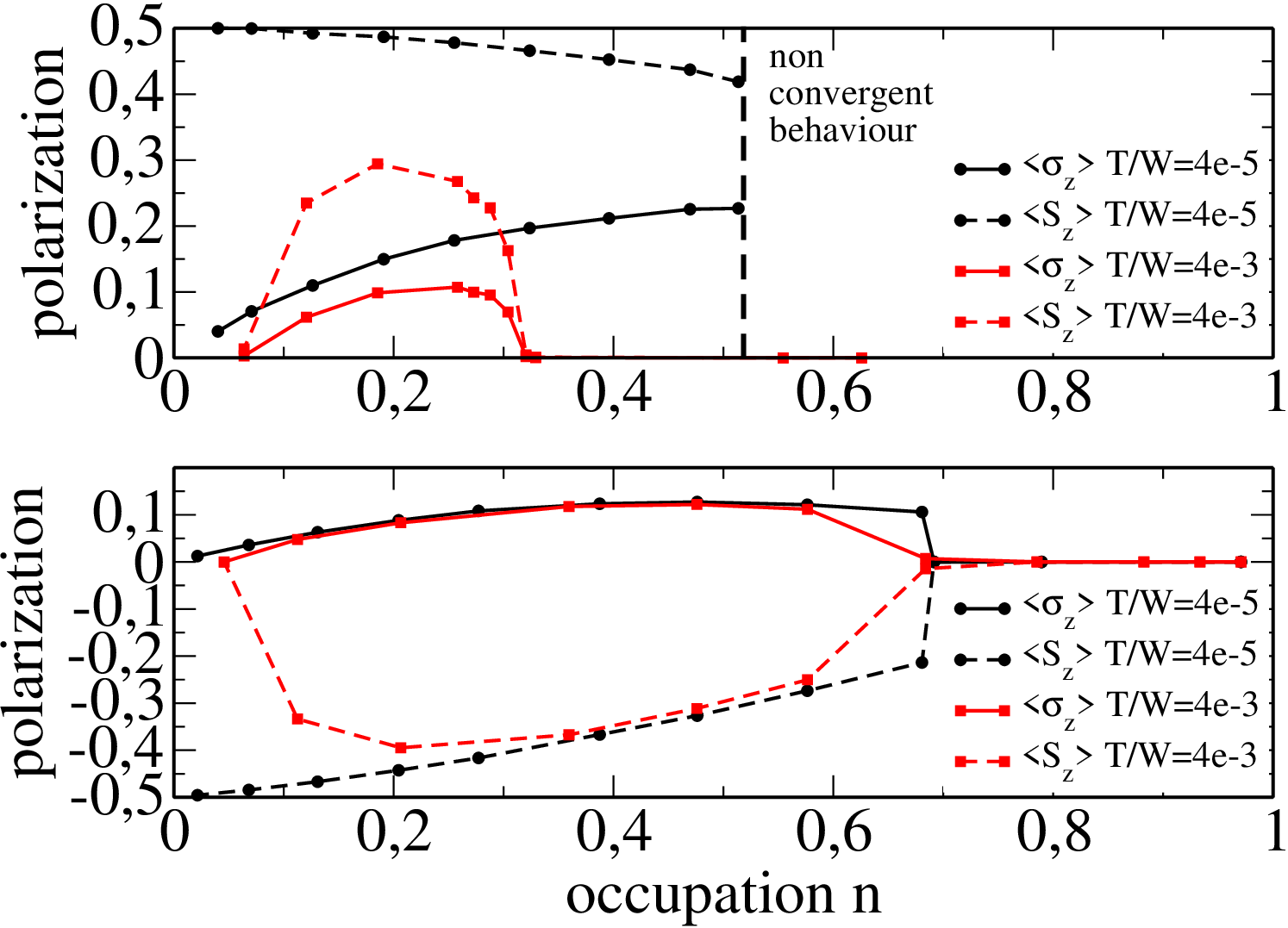}
\end{center}
\caption[]{(color online) Polarization of the ferromagnetic solutions. The upper (lower) panel
  corresponds to ferromagnetic (antiferromagnetic) coupling. Other
parameters as in Fig.\ \ref{Phase0101}.}\label{ferro01} 
\end{figure}
The dependence of the polarization on $n$ and $T$ in the ferromagnetic phase for the
present values of $U$ and $|J|$ is shown in Fig.\ \ref{ferro01} 
for $J>0$ (upper panel) and $J<0$ (lower panel). As in Fig.\ \ref{Ferro001},
full lines represent $\langle s_z\rangle$ and dashed lines $\langle S_z\rangle$.
For high temperatures (squares) we again find second order phase transitions,
indicated by the smooth simultaneous vanishing of both $\langle s_z\rangle$
and $\langle S_z\rangle$, while for low $T$ (circles) and $J>0$ the peculiar
behavior at the phase boundary to the ``incommensurate'' phase already seen in
Fig.\ \ref{Ferro001} is observed.

A markable difference to the behavior for $U=0$ collected in Fig.\ \ref{Ferro001}
can be observed for antiferromagnetic coupling $J<0$. Here, the polarization
now vanishes discontinously at $n_c^{(FM)}$ for $T\to0$, indicating a first
order phase transition in contrast to the second order type observed for $U=0$.
Thus, while the results at $U=0$ would predict a quantum critical point
at $n_c^{(FM)}$, such a scenario is obvioulsy destroyed by Coulomb correlations
in the conduction band, at least within DMFT.

Note that for the bare Hubbard model no ferromagnetic
phase exists in this parameter regime for a bipartite lattice.
It is thus clearly the spin-electron interaction which leads to
ferromagnetism. Our results are consistent with earlier studies by other
groups, who however treated the electron-electron interaction in an effective
medium approach,\cite{nolting:03,kienert:03,golosov:05,stier:07} while we are
able to treat the on-site correlations exactly within DMFT and NRG. Especially
our Curie temperatures are located in the same range as in these earlier studies.

\begin{figure}[htb]
\begin{center}
\includegraphics[width=0.45\textwidth,clip]{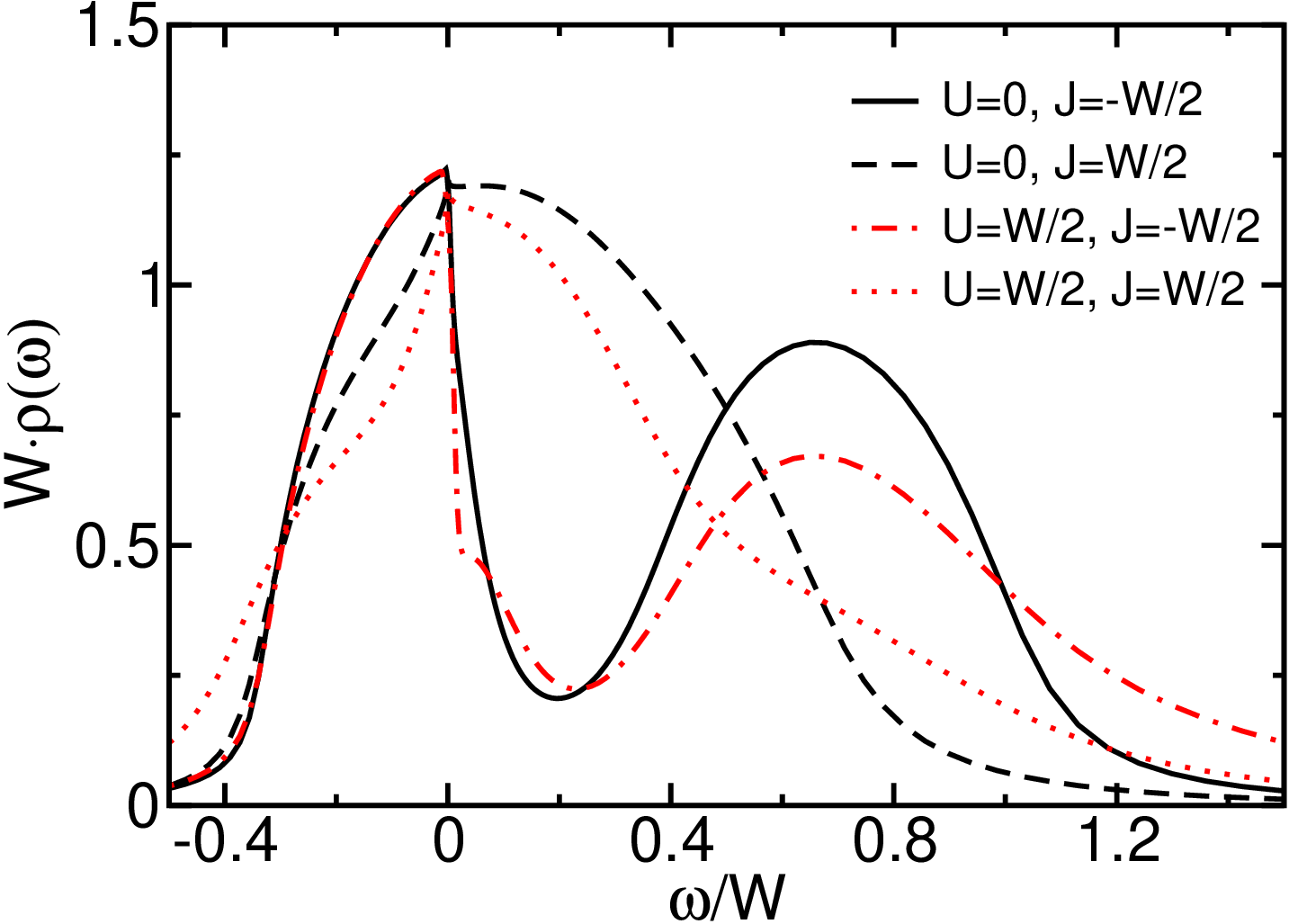}
\end{center}
\caption[]{(color online) DOS for $n=0.6$, $T=4\cdot10^{-5}$,
  $U=0$, $W/2$ and $J=\pm W/2$.}\label{DOS_n06} 
\end{figure}

A possible reason for the tendency of both finite $U$ and negative $J$ to
stabilize ferromagnetism is the strong asymmetry induced by $U$ and $J$\cite{tasaki:98,ulmke:98}.
As can be seen from Fig.\ \ref{DOS_n06}, this effect is most pronounced
for $J<0$. For $J>0$, a weaker, but nonetheless significant redistribution
of spectral weight is observed, which can explain the increase of $n_c^{(FM)}$
here, too.
\section{Summary}
In this paper we discussed the magnetic properties of the 
extended Kondo lattice model, where in addition to an exchange coupling between
itinerant states and localized spins the band electrons are subject to a
local Coulomb interaction. Such a model is expected to describe the properties
of e.g.\ transition metal compounds, where part of the $d$-states are localized
in the crystalline environment due Coulomb correlations and form a local spin
which couples to the remaining, usually itinerant $d$-states either due to Hund's exchange
or hybridization.

These materials typically have a rich magnetic phase diagram, and it is an 
interesting question, how these two interactions cooperate or compete in
driving different magnetic phases. A particularly interesting question in
this respect is, to what extent, qualitatively and quantitatively, the concept of
the RKKY approximation of an effective interaction between the local spins
mediated by the itinerant conduction states holds.

We treat the model within DMFT, using NRG as solver for the effective impurity
problem. The calculations were done for a bipartite lattice, allowing for
homogeneous ferromagnetism and N\'eel antiferromagnetism. At half filling,
we find an antiferromagnetic phase, as expected, for all ferromagnetic
Kondo couplings $J>0$. For antiferromagnetic coupling $J<0$, there exists
a critical $J_c$, where the system becomes a paramagnetic Kondo insulator.\cite{doniach:77}

For $J>J_c$, this antiferromagnetic phase prevails for finite doping up to
a critical filling $n_c^{(AF)}$ depending on $J$ and the local Coulomb interaction
$U$. For finite $U$ and $J>0$, we in addition find phase separation between the
antiferromagnetic insulator with $n=1$ and an antiferromagnetic metal
with $n<1$. For $U=0$ we find no convincing evidence that such a phase separation
exists, too, but the numerical results are not sufficiently clear to rule it out
either.
For $J<J_c$ no antiferromagnetic phase exists at all.

Below a second critical filling $n_c^{(FM)}$ a homogeneous ferromagnet is found.
Quite interestingly, the extent and stability of this phase is increased
by both a finite $U$ and an antiferromagnetic $J$. The reason for
this stabilization can quite likely be traced back to the introduction
of a strong asymmetric redistribution of spectral weight in the electronic spectral function. An interesting aspect in connection with quantum
critical points in e.g.\ ferromagnetic heavy fermion compounds is that
for antiferromagnetic Kondo exchange a finite Coulomb correlation among the
conduction states has the tendency to change the order of the phase transition
at $n_c^{(FM)}$ from second order without correlation to first order with correlations.

 We find that for
ferromagnetic Kondo coupling $J$ the predictions by
RKKY can be used to at least qualitatively account for the different phases.
For moderate and large $J<0$, however, RKKY cannot even qualitatively predict
the phases correctly, underestimating the ferromagnetic phase grossly. Moreover,
the variations of the polarizations of the local spin and bandelectrons 
seem to follow RKKY for very small $|J|$, but already deviate substantially
for moderate values.
\begin{acknowledgments}
We want to thank helpful discussions with Andreas Honecker, Akihisa
Koga and Dieter Vollhardt. This work was supported by the DFG through
PR298/10. Computer support was provided by the Gesellschaft f\"ur
wissenschaftliche Datenverarbeitung in G\"ottingen and the
Norddeutsche Verbund f\"ur Hoch- und H\"ochstleistungsrechnen.
\end{acknowledgments}

\end{document}